\documentclass{pasj00}

\begin{document}
\SetRunningHead{Hirota et al.}{VERA astrometry of SVS~13 in NGC~1333}
\Received{2007/06/06}
\Accepted{2007/09/10}

\title{Astrometry of Water Maser Sources in Nearby Molecular Clouds with VERA 
--- II. SVS~13 in NGC~1333}
\author{
Tomoya \textsc{Hirota},\altaffilmark{1,2}
Takeshi \textsc{Bushimata},\altaffilmark{1,3}
Yoon Kyung \textsc{Choi},\altaffilmark{1,4}
Mareki \textsc{Honma},\altaffilmark{1,2} 
Hiroshi \textsc{Imai},\altaffilmark{5} \\
Kenzaburo \textsc{Iwadate},\altaffilmark{6}
Takaaki \textsc{Jike},\altaffilmark{6} 
Osamu \textsc{Kameya},\altaffilmark{2,6} 
Ryuichi \textsc{Kamohara},\altaffilmark{1} 
Yukitoshi \textsc{Kan-ya},\altaffilmark{7} \\
Noriyuki \textsc{Kawaguchi},\altaffilmark{1,2,3}
Masachika \textsc{Kijima},\altaffilmark{2} 
Hideyuki \textsc{Kobayashi},\altaffilmark{1,3,4,6}
Seisuke \textsc{Kuji},\altaffilmark{6} \\
Tomoharu \textsc{Kurayama},\altaffilmark{1}
Seiji \textsc{Manabe},\altaffilmark{2,6} 
Takeshi \textsc{Miyaji},\altaffilmark{1,3}
Takumi \textsc{Nagayama},\altaffilmark{8} 
Akiharu \textsc{Nakagawa},\altaffilmark{5} \\ 
Chung Sik \textsc{Oh},\altaffilmark{1,4}
Toshihiro \textsc{Omodaka},\altaffilmark{5}
Tomoaki \textsc{Oyama},\altaffilmark{1} 
Satoshi \textsc{Sakai},\altaffilmark{6} 
Tetsuo \textsc{Sasao},\altaffilmark{9,10} \\
Katsuhisa \textsc{Sato},\altaffilmark{6} 
Katsunori M. \textsc{Shibata},\altaffilmark{1,2,3} 
Yoshiaki \textsc{Tamura},\altaffilmark{2,6} 
and Kazuyoshi \textsc{Yamashita}\altaffilmark{2}
}

\altaffiltext{1}{Mizusawa VERA Observatory, National Astronomical Observatory of Japan, \\
  2-21-1 Osawa, Mitaka, Tokyo 181-8588}
\altaffiltext{2}{Department of Astronomical Sciences, Graduate University for Advanced Studies, \\
  2-21-1 Osawa, Mitaka, Tokyo 181-8588}
\altaffiltext{3}{Space VLBI Project, National Astronomical Observatory of Japan, \\
  2-21-1 Osawa, Mitaka, Tokyo 181-8588}
\altaffiltext{4}{Department of Astronomy, Graduate School of Science, The University of Tokyo, \\
  7-3-1 Hongo, Bunkyo-ku, Tokyo 113-0033}
\altaffiltext{5}{Faculty of Science, Kagoshima University, \\
  1-21-35 Korimoto, Kagoshima, Kagoshima 890-0065}
\altaffiltext{6}{Mizusawa VERA Observatory, National Astronomical Observatory of Japan, \\
  2-12 Hoshi-ga-oka, Mizusawa-ku, Oshu-shi, Iwate 023-0861}
\altaffiltext{7}{Department of Astronomy, Yonsei University, \\
  134 Shinchong-dong, Seodaemun-gu, Seoul 120-749, Republic of Korea}
\altaffiltext{8}{Graduate School of Science and Engineering, Kagoshima University, \\
  1-21-35 Korimoto, Kagoshima, Kagoshima 890-0065}
\altaffiltext{9}{Department of Space Survey and Information Technology, Ajou University, \\
  Suwon 443-749, Republic of Korea}
\altaffiltext{10}{Korean VLBI Network, Korea Astronomy and Space Science Institute, \\
  P.O.Box 88, Yonsei University, 134 Shinchon-dong, Seodaemun-gu, Seoul 120-749, Republic of Korea}
\email{tomoya.hirota@nao.ac.jp}

\maketitle

\begin{abstract}
We report on the results of multi-epoch VLBI observations with VERA (VLBI Exploration of 
Radio Astrometry) of the 22~GHz H$_{2}$O masers associated with the young stellar object 
SVS~13 in the NGC~1333 region. We have carried out phase-referencing VLBI astrometry 
and measured an annual parallax of the maser features 
in SVS~13 of 4.25$\pm$0.32~mas, corresponding to the distance of 235$\pm$18 pc 
from the Sun.  Our result is consistent with a photometric distance of 220~pc 
previously reported. 
Even though the maser features were detectable only for 6~months, 
the present results provide the distance to NGC~1333 with 
much higher accuracy than photometric methods. 
The absolute positions and proper motions have been derived, revealing that 
the H$_{2}$O masers with the LSR (local standard of rest) velocities of 7-8~km~s$^{-1}$ 
are most likely associated with VLA4A, which is a radio counterpart of SVS~13. 
The origin of the observed proper motions of the maser features are currently difficult to 
attribute to either the jet or the rotating circumstellar 
disk associated with VLA4A, which should be investigated through future high-resolution 
astrometric observations of VLA4A and other radio sources in NGC~1333. 
\end{abstract}

\section{Introduction}

The reflection nebula NGC~1333 is one of the nearest low-mass 
star-forming regions and is associated with the classical Herbig-Haro (HH) objects 
HH~7-11 (\cite{herbig1974}; \cite{strom1974}). 
Despite the importance of NGC~1333 for studying star-formation processes 
and the overall structure of the larger molecular cloud complex,
the distance to NGC~1333 is still quite uncertain. 
NGC~1333 is located in the Perseus complex of dark clouds, 
which forms a chain of several clouds with a size of $7^{\circ} \times 3^{\circ}$, 
elongated perpendicular to the Galactic plane (\cite{cernis1990}, 1993). 
The molecular clouds associated with NGC~1333 are 
located at the western edge of the chain, while the
young open clusters IC~348 and Perseus~OB2 are at the eastern edge of 
this chain (\cite{cernis1990}, 1993). 

The large scale structure of the Perseus dark cloud complex has been studied extensively
(\cite{cernis1990}, 1993; \cite{dezeeuw1999}). 
According to these studies, there is a gradient in distance across the complex 
(see Fig.9 of \cite{cernis1993});
NGC~1333 is the nearest cloud at a distance of 220~pc (\cite{cernis1990}) 
and IC~348 and the Perseus OB2 association are at a distance of 318-340~pc 
(\cite{cernis1993}; \cite{dezeeuw1999}). 
Although Hipparcos provided an accurate distance to 
the Perseus OB2 association, based on an annual parallax measurement 
of 318$\pm$27~pc (\cite{dezeeuw1999}), 
the distance to NGC~1333 and the other dark clouds have been measured 
only through optical photometry, with a typical uncertainties of about 25\% 
(\cite{cernis1990}, 1993). 
Therefore, further precise astrometry allowing an accurate 
annual parallax measurement is necessary to reveal overall structure of 
the chain of the molecular clouds in the Perseus region. 

The young stellar object (YSO) SVS~13 is located at the base of HH~7-11 in NGC~1333 
and has been proposed to be a powering source of the jets and outflows in this region 
(\cite{strom1976}).  High-resolution 
observations with radio interferometers at centimeter and millimeter 
wavelengths have been conducted to investigate the complex nature of this system. 
Very Large Array (VLA) observations at
3.6~cm wavelength reveal a radio continuum source VLA4 that appears to be 
associated with SVS~13 (\cite{rodriguez1997}, 1999). 
Higher resolution VLA observations of SVS~13 at the same wavelength 
resolved VLA4 into a double radio source, denoted 
VLA4A and VLA4B (\cite{anglada2000}), with the separation angle of 0.3\arcsec. 
\citet{anglada2000} reported that the position of SVS~13 coincides with 
the western component, VLA4A, while that of the millimeter source (\cite{looney2000}) 
is associated with the eastern component, VLA4B.  VLA4B is suggested to have larger 
amount of circumstellar material than VLA4A, based on the subsequent high-resolution 
VLA observations in the 7~mm band (\cite{anglada2004}). 

Although HH~7-11 have been proposed to be powered by SVS~13, 
\citet{rodriguez1997} argued that another radio continuum source VLA3, 
located  6\arcsec \ southwest of SVS~13, is a more favorable candidate for 
the powering source due to the better alignment of VLA3 with the HH objects. 
In contrast, \citet{bachiller2000} presented the results of 
the interferometric CO $J$=2-1 observations of HH~7-11 that indicate that 
the extremely high velocity (EHV) molecular outflow is powered by SVS~13, 
rather than VLA3 which has no high velocity CO outflow. 
\citet{looney2000} also suggested that VLA3 (or millimeter source A2 
in their paper) is a less likely candidate for the source that powers HH~7-11. 
\citet{rodriguez2002} claimed that both VLA3 and SVS~13 (VLA4B is favorable rather 
than 4A) could be the powering source of HH~7-11. 
Despite the fact that the axes of the HH jets and molecular outflows are similar, 
their exciting sources could be different, since there are several YSOs in 
a small area. In fact, it is possible that there is more than one molecular 
outflow because, as noted by \citet{rodriguez2002}, inspection of Figure 2 of 
\citet{bachiller2000} shows that the EHV outflow is associated with VLA4, while 
the standard high velocity (SHV) gas is found in the vicinity of both VLA3 and VLA4. 
Thus the origin of the jets, molecular outflows, and HH~7-11 is still ambiguous
and might be revealed through accurate proper motion measurements. 

Proper motions can be accurately measured by phase-referencing VLBI observations. 
If one employs extragalactic radio sources as the position references, 
one can measure the absolute position of the target source (\cite{beasley1995}), 
making it feasible to derive its annual parallax as well as the absolute proper motion. 
Highly precise VLBI astrometric observations 
have been carried out for the Galactic CH$_{3}$OH and H$_{2}$O maser source W3(OH) with 
the NRAO Very Long Baseline Array (VLBA), which yielded distances of 
2.0~kpc from the Sun with uncertainties of about 2\%, based on the annual parallax method 
(\cite{xu2006}; \cite{hachisuka2006}). 

Recently, we have constructed a new VLBI network in Japan called VERA, 
VLBI Exploration of Radio Astrometry (\cite{kobayashi2003}), 
which is the first VLBI array dedicated to phase-referencing astrometry. 
Initial results of VERA observations have been reported 
(e.g. \cite{sato2007}; \cite{honma2007}; \cite{hirota2007}; \cite{imai2007}), 
in which absolute proper motions and annual parallaxes are measured 
for galactic H$_{2}$O maser sources at distances 
ranging from 178~pc for IRAS~16293-2422 (\cite{imai2007}) 
to 5.28~kpc for S269 (\cite{honma2007}), 
demonstrating VERA's astrometric capability.

In this paper, we present the results of astrometry of 
the H$_{2}$O masers associated with SVS~13 in NGC~1333 with VERA. 
These observations have been done as one of the initial 
scientific projects of VERA: measurements of annual parallaxes of 
nearby molecular clouds (e.g. \cite{hirota2007}; \cite{imai2007}). 
SVS~13 is known to be one of the brightest H$_{2}$O maser sources 
among the known low-mass YSOs (\cite{claussen1996}; \cite{furuya2003}). 
In addition, there is a bright ICRF source, J0336$+$3218, which is separated 
from SVS~13 by an angle of $1.89^{\circ}$ (\cite{fey2004}).
The aim of this paper is to measure the distance to NGC~1333 
as well as to explore the powering source of the jets and outflows in this region 
through the maser positions and proper motions.

\section{Observations}

The VERA observations of H$_{2}$O masers ($6_{1 6}$-$5_{2 3}$, 22235.080 MHz) 
associated with SVS~13 were carried out for 10~hours 
in each of 7 observing sessions from Nov. 2004 to May 2005 
(2004/317, 2004/355, 2005/021, 2005/052, 2005/080, 2005/113, and 2005/140; 
hereafter an observing session is denoted by year/day of the year) 
at intervals of about 1~month. 
All 4 stations of VERA (see Fig.1 of \cite{petrov2007}) took part in all sessions, 
providing a maximum baseline length of 2270~km. 

Observations were made in the dual beam mode; the H$_{2}$O masers associated 
with SVS~13 and a reference source J0336$+$3218 (\cite{fey2004}), 
with the separation angle of $1.89^{\circ}$, were observed simultaneously. 
The instrumental phase difference between the two beams was measured 
continuously during the observations by injecting 
artificial noise sources into both beams at each station 
(\cite{kawaguchi2000}; \cite{honma2003}). 
The typical value of the phase drift between the two beams was 3 degrees per hour, 
which was removed from the data.

Left-handed circular polarization was received and sampled with 2-bit 
quantization and filtered using the VERA digital filter unit (\cite{iguchi2005}). 
The data were recorded onto magnetic tapes at a rate of 128~Mbps, 
with two IF channels of 16~MHz bandwidth each for both SVS~13 and J0336$+$3218. 
J0336$+$3218 was detected with a peak intensity greater than 
1.2~Jy~beam$^{-1}$ in all the sessions, so it was also used for bandpass and delay calibration. 
System temperatures and atmospheric attenuation were measured with 
the chopper-wheel method (\cite{ulich1976}), and typical values were 100-400~K, 
depending on weather conditions and elevation angles of the observed sources. 
The aperture efficiencies of the antennas ranged from 45 to 52\%. 
Correlation processing was carried out on the Mitaka FX correlator 
(\cite{chikada1991}) located at the NAOJ Mitaka campus. 
For H$_{2}$O maser lines, the spectral resolution was set to be 15.625~kHz, 
corresponding to the velocity resolution of 0.21~km~s$^{-1}$. 

\section{Data Reduction}

\begin{figure}
  \begin{center}
    \FigureFile(80mm,80mm){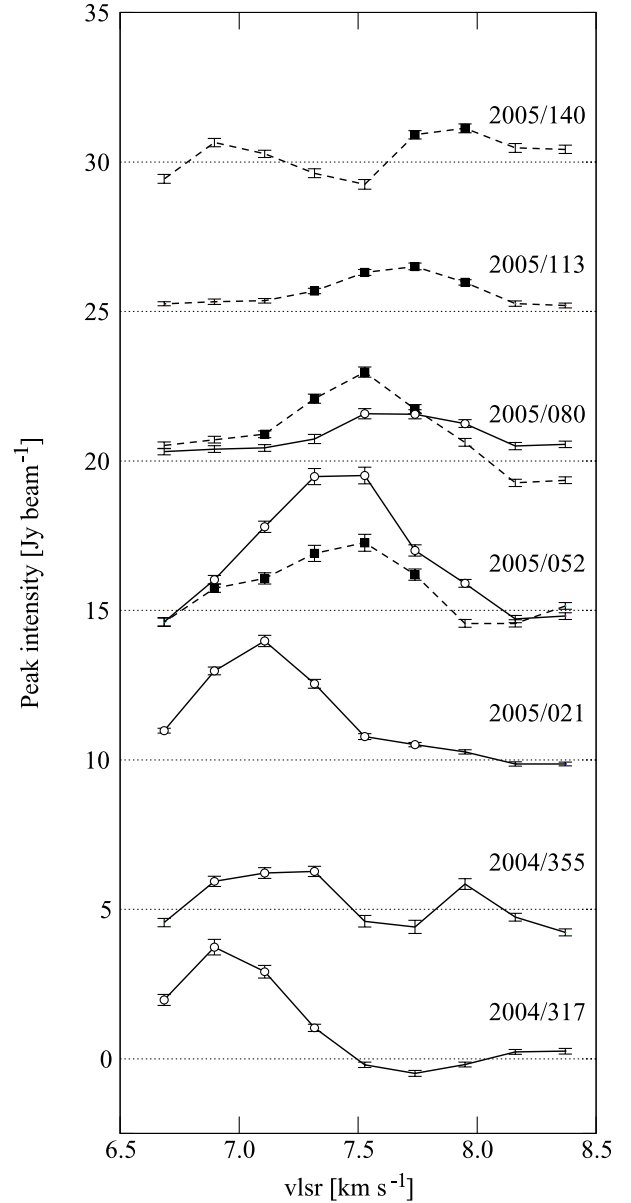}
  \caption{Spectra of the H$_{2}$O maser features associated with SVS~13. 
  Solid and dashed lines represent the features 1 and 2, respectively. 
  Error bars represent the fitting errors (1$\sigma$) given by the AIPS task JMFIT. 
  Maser spots with signal to noise ratios larger than 5 times the noise  
  (5$\sigma$) for at least two consecutive channels are considered to be significant 
  detections and are plotted with 
  open circles for feature 1 and filled squares for feature 2 (see text). 
  For lower signal to noise (non-detections) we only plot error bars. 
  Note that negative intensities are expected due to thermal noise variations. 
  Spectra at different epochs are offset by 5~Jy~beam$^{-1}$ steps 
  to clarify the time variations of the spectra. }
  \label{fig-spectra}
  \end{center}
\end{figure}

Data reduction was performed using 
the NRAO Astronomical Image Processing System (AIPS). 
Amplitude and bandpass calibration were done for the target 
(SVS~13) and reference (J0336$+$3218) sources independently. 
For phase calibration, fringe fitting was done with the AIPS task FRING on 
the phase reference source (J0336$+$3218), and 
the solutions were applied to the target source (SVS~13). 
In addition, we applied the results of the dual-beam phase 
calibration measurements (\cite{kawaguchi2000}).  We also 
corrected for the approximate delay model adopted in the correlation processing 
(\cite{honma2007}) and for drifts 
of the visibility phase caused by the Earth's atmosphere based on the GPS measurements. 

We first searched for H$_{2}$O maser ``spots'', 
defined as emission occurring in a single velocity channel, 
in fringe-rate maps with the AIPS task FRMAP. 
We found that the H$_{2}$O masers were detected toward two different sources in NGC~1333; 
one was associated with SVS~13, which was previously identified to be H$_{2}$O(A) by 
\citet{haschick1980}, 
and another was located 34\arcsec \ southwest from SVS~13, corresponding to the radio continuum 
source VLA2 and H$_{2}$O(B) by \citet{haschick1980}. 
The H$_{2}$O masers associated with VLA2 were detected only in the first three 
sessions from 2004/317 to 2005/021. 
Because the positions of the maser spots associated with VLA2 are 
shifted by 34\arcsec \ southwest of the phase-tracking center, 
the synthesized images are significantly distorted due 
to the time-averaging smearing effect (\cite{cotton1999}), 
making it almost impossible to conduct accurate astrometry. 
Therefore, we do not discuss the masers associated with VLA2 in this paper. 

High resolution synthesis imaging and deconvolution (CLEAN) were done 
for the maser spots associated with SVS~13 using the AIPS task IMAGR. 
The image size of each map was 51.2~mas$\times$51.2~mas 
with the pixel size of 0.1~mas$\times$0.1~mas. 
The naturally-weighted synthesized-beam size (FWMH) was typically 
1.3~mas$\times$0.9~mas with a position angle of $-50^{\circ}$. 
With a net integration time of 7~hours for the H$_{2}$O maser lines, 
the resultant rms noise levels of the phase-referenced images ranged from 
0.06~Jy~beam$^{-1}$ to 0.16~Jy~beam$^{-1}$ in a single spectral channel. 
We regard the H$_{2}$O maser spot to be real if the signal to noise ratio of 
the peak intensity is larger than 5 times 
the noise level (5$\sigma$) for at least two consecutive channels, provided
the positions are coincident within the synthesized beam size. 
In order to improve the signal to noise ratios in the phase-referenced images, 
we made integrated intensity maps of the maser features, 
by summing the channel maps over the detected velocity range. 
The peak positions and peak intensities 
of maser features were derived by fitting elliptical Gaussian 
brightness distributions to  
these integrated intensity maps using the AIPS task JMFIT. 
The formal uncertainties in the feature positions given by JMFIT were 
0.03-0.1~mas, depending on the signal to noise ratios of the features and 
possibly their spatial structure. 
The uncertainties in the peak intensities were 0.1-0.3~Jy~beam$^{-1}$ 
(1$\sigma$) which were approximately equal to or slightly larger than 
the rms noise levels of the phase-referenced images. 

\section{Results}
\subsection{Structure of the H$_{2}$O maser features}

\begin{figure}
  \begin{center}
    \FigureFile(80m,80mm){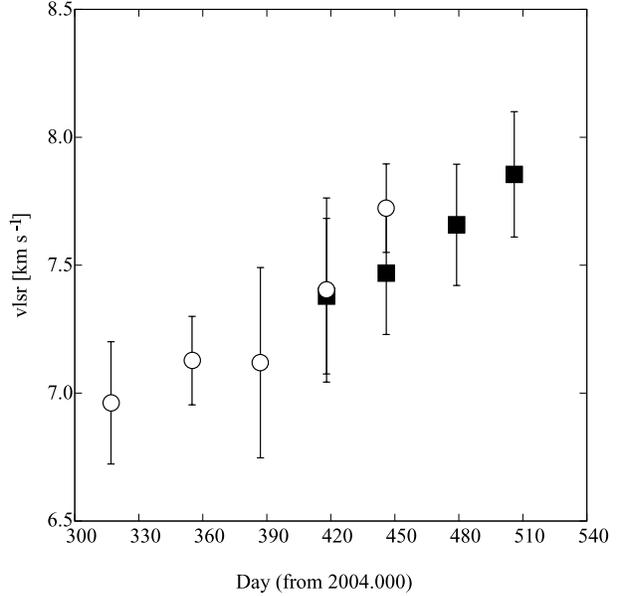}
  \caption{Peak LSR velocities of the H$_{2}$O maser features. Open circles and filled squares 
   represent the peak velocities for the features 1 and 2, respectively, which are derived 
   from the weighted means of the spectra shown in Figure \ref{fig-spectra}. 
   Error bars indicate the standard deviations (1$\sigma$) of the velocities.
   }
  \label{fig-vshift}
  \end{center}
\end{figure}

Figure \ref{fig-spectra} shows the spectra of the H$_{2}$O masers associated with SVS~13, 
which are obtained by fitting Gaussian brightness distributions to each channel map. 
We detected two spatially distinct H$_{2}$O maser features at the LSR velocities 
of 7-8~km~s$^{-1}$, 
which agree well with the ambient cloud velocity (e.g. \cite{rodriguez2002}). 
Hereafter we call them as feature 1 and 2, as explained later. 
We did not find other velocity components detected previously 
by \citet{haschick1980}, \citet{wootten2002}, and \citet{rodriguez2002}, due to 
the variability of the H$_{2}$O masers (e.g. \cite{claussen1996}; \cite{furuya2003}). 
As shown in Figure \ref{fig-vshift}, 
the peak velocity of each maser feature was found to drift systematically 
from 7.0~km~s$^{-1}$ to 7.9~km~s$^{-1}$ during the observing period of 6~months, 
at a rate of 1.7~km~s$^{-1}$~yr$^{-1}$. 
It is likely that the observed velocity drift indicates true acceleration of 
the maser feature as discussed later, although we cannot rule out 
the possibilities of a change in strength of blended hyperfine components 
(\cite{walker1984}) or the structure in the maser feature. 

\begin{figure}
  \begin{center}
    \FigureFile(80mm,80mm){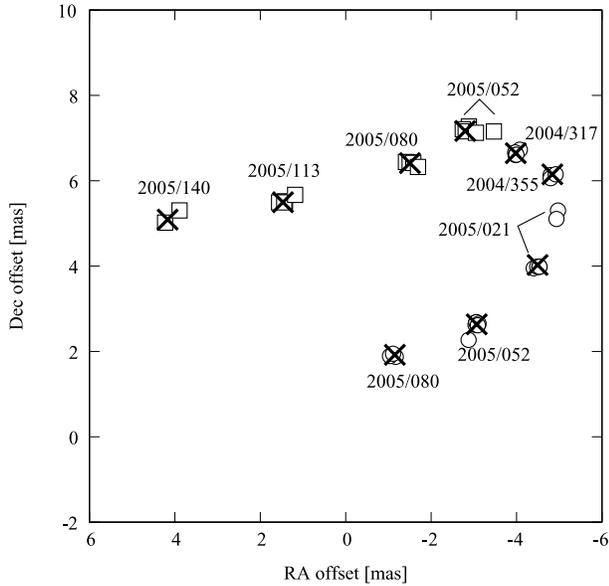}
  \caption{Distribution of H$_{2}$O maser spots associated with SVS~13. 
   Open circles and squares represent the positions of the maser spots 
   belonging to the features 1 and 2, respectively, while 
   crosses represent features that are the collections 
   of spots integrated over contiguous spectral channels. 
   Note that the formal errors given by the Gaussian fitting with 
   the AIPS task JMFIT are as small as 0.1~mas or less, which cannot be 
   shown in the figure. 
   The positions are with respect to the reference position at 
   $\alpha(J2000)=$03h29m03.72465s, $\delta(J2000)=+31$d16\arcmin03.8015\arcsec. }
  \label{fig-allspot}
  \end{center}
\end{figure}

Figure \ref{fig-allspot} shows the positions of the maser spots and features 
associated with SVS~13. 
Initially, we found only feature 1 in the sessions 2004/317 - 2005/021, 
as seen in the spectra (Figure \ref{fig-spectra}). 
In the session 2005/052, feature 2 appeared at 4.5 mas north of feature 1. 
The two features existed at almost the same LSR velocities of about 7.5~km~s$^{-1}$ 
in the sessions 2005/052 and 2005/080. 
In 2005/080, feature~2 became brighter than feature~1, 
which disappeared in the subsequent session 2005/113. 
Feature~2 remained until the session 2005/140, while all the H$_{2}$O masers 
associated with SVS~13 finally disappeared after the 6-month monitoring period of our observations. 

We can see the internal structures of the maser features, 
which are the collections of several maser spots, in Figure \ref{fig-allspot}. 
The dispersions of the maser spots within the features, 0.04-0.27~mas, 
are significantly smaller than the synthesized beam size, except 
for those detected in the session 2005/021 in which we found two 
different groups of spots with the separation of 1~mas. 
These northern spots correspond to the weak red shifted shoulder 
at the LSR velocity of 7.5-7.7~km~s$^{-1}$ as seen 
in Figure \ref{fig-spectra} and can be recognized as part of the elongated structure 
of the maser feature in the integrated intensity map. 
These results suggest that the peak position of the maser feature 
possibly affect the astrometric accuracy (e.g. \cite{hirota2007}; 
\cite{imai2007}), as discussed in the next section. 

\subsection{Astrometry of the H$_{2}$O maser features}

\begin{figure*}[th]
  \begin{center}
    \FigureFile(160mm,160mm){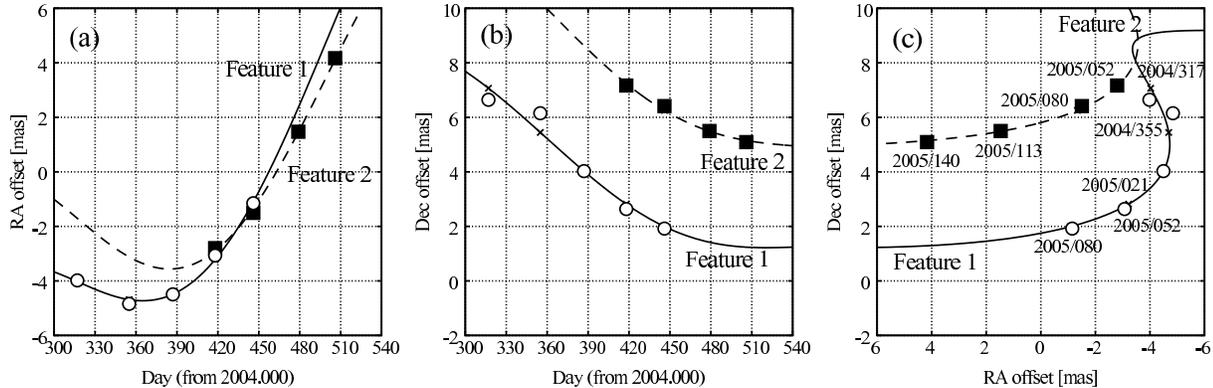}
  \caption{Position measurements of the maser features associated with SVS~13. 
   (a) The movement of the maser features in right ascension as a function of time. 
   (b) The same as (a) in declination. 
   (c) The movement of the maser features on the sky. Solid and dashed lines represent the 
   best fit model with the annual parallax and linear proper motion for the 
   maser features 1 and 2, respectively. 
   Open circles and filled squares represent the observed positions 
   of the maser features 1 and 2, respectively, and small crosses represent 
   predicted positions of the maser features. 
   Note that the formal errors given by the Gaussian fitting with 
   the AIPS task JMFIT are as small as 0.1~mas or less, which cannot be 
   shown in the figure. 
   The reference position is the same as in Figure \ref{fig-allspot}. }
  \label{fig-ra}
  \end{center}
\end{figure*}

As shown in Figure \ref{fig-allspot}, the movement of the masers significantly 
deviates from a 
simple linear motion; as we will show the data can be well modeled by the 
effects of an annual parallax. 
Assuming that the movement of the maser feature is the sum of a
linear motion and the annual parallax, we derive the proper motions in right ascension 
$\mu_{\alpha} \cos \delta$ and declination $\mu_{\delta}$, the initial positions in right ascension $\alpha_{0}$
and declination $\delta_{0}$, and the annual parallax $\pi$ for the maser feature by 
a least-squares analysis as summarized in Table \ref{tab-results} and Figure \ref{fig-ra}. 
In the least-squares analysis, 
we determined the proper motions and initial positions of two maser features 
independently, while the annual parallax of SVS~13 is common for both features. 
According to \citet{honma2007} and \citet{hirota2007}, 
the precision of the derived annual parallax is significantly improved when we 
use only the data for right ascension, as this data is less affected by 
atmospheric modeling errors.  
In this manner, we find the annual parallax of the H$_{2}$O maser features 
associated with SVS~13 in NGC~1333 to be 4.25$\pm$0.32~mas, corresponding 
to a distance of 235$\pm$18~pc.
We note that the parallax derived from the right ascension data fits the declination
data quite well.

\begin{table}[tb]
\begin{center}
\caption{Results of the least-squares analysis for the annual parallax and absolute proper motion measurements}
\label{tab-results}
\begin{tabular}{lrr}
 \hline\hline
Parameter                         & Feature 1 & Feature 2\\ 
\hline
$\alpha_{0}$ (mas)$^{a}$          & -4.08(0.14)          &   1.17(0.36)    \\
$\delta_{0}$ (mas)$^{a}$          &  6.14(0.31)          &   8.38(0.32)    \\
$\mu_{\alpha} \cos \delta$ (mas~yr$^{-1}$)  & 17.9(0.9)  &  10.6(1.7)      \\
$\mu_{\delta}$ (mas~yr$^{-1}$)    &  -7.9(1.4)           & -10.0(2.1)      \\
$\mu$ (mas~yr$^{-1}$)             &  19.6(1.0)           &  14.6(1.9)      \\
$v_{t}$ (km~s$^{-1}$)             &  21.8(1.1)           &  16.3(2.1)      \\
PA (degrees)                      & 114                  & 133            \\
$\pi$ (mas)                       & \multicolumn{2}{c}{4.25(0.32)}        \\
$\sigma_{\alpha}$ (mas)           & \multicolumn{2}{c}{0.10}              \\
$\sigma_{\delta}$ (mas)           & \multicolumn{2}{c}{0.29}              \\
\hline \\
\multicolumn{3}{l}{Note --- Numbers in parenthesis represent the } \\
\multicolumn{3}{l}{\quad estimated uncertainties. Annual parallax $\pi$ is } \\
\multicolumn{3}{l}{\quad derived from the right ascension data only. } \\
\multicolumn{3}{l}{\quad $\sigma_{\alpha}$ and $\sigma_{\delta}$ are the rms deviations } \\
\multicolumn{3}{l}{\quad of the post-fit residuals in the right ascension  } \\
\multicolumn{3}{l}{\quad and declination directions. } \\
\multicolumn{3}{l}{$a$: The position offsets are measured with respect }\\ 
\multicolumn{3}{l}{\quad to the reference position. }\\
\end{tabular}
\end{center}
\end{table}

According to our astrometric results, 
the standard deviations of the post-fit residuals  
are 0.10~mas and 0.29~mas in right ascension and declination, 
respectively, as listed in Table \ref{tab-results}. 
Were we to assume distances of 300~pc and 350~pc from the Sun (similar to
that of IC 348 and the Perseus OB2 association), 
the standard deviation in right ascension degrades to
0.19~mas and 0.26~mas, respectively. Thus, the distance to NGC~1333 
is most likely 235~pc, which is consistent with the photometric 
distance to NGC~1333 reported by \citet{cernis1990} of 220~pc with an 
uncertainty of 25\%, rather than the larger value of about 300~pc 
(e.g. \cite{herbig1983}; \cite{dezeeuw1999}). 
This is the first time that the distance to NGC~1333 is determined 
based on an annual parallax measurement with an uncertainty of only 8\%. 
Even though the observing period was 
as short as 6~months, our results provide a strong constraint on the distance 
to NGC~1333 with higher precision than that of the photometric method 
(\cite{cernis1990}). 

Unfortunately, the masers associated with SVS~13 completely disappeared after the 6-month 
monitoring observations with VERA. We will continue long-term monitoring 
observations of the H$_{2}$O maser sources in NGC~1333, including SVS~13 and 
others such as IRAS2 and IRAS4 (\cite{rodriguez2002}), which may enable us to improve 
the accuracy of the annual parallax measurements with VERA. 

Combining our new results with those of \citet{dezeeuw1999} based on 
the Hipparcos measurements of the distance to the Perseus OB2 association, 
318$\pm$27~pc, we provide a definite evidence for a distance gradient along 
the chain of molecular clouds in the Perseus region as proposed by 
\citet{cernis1990} and \citet{cernis1993}. 
Although we determine the distance to only one H$_{2}$O maser source 
associated with SVS~13 in NGC~1333 in this paper, 
we should be able to reveal the overall 3-dimensional structure of 
the Perseus molecular cloud complex 
through the further VLBI astrometry of other H$_{2}$O maser sources in this region. 

\section{Discussions}
\subsection{Error sources in our astrometry}

The standard deviations of the post-fit residuals from the least-squares analysis, 
0.10~mas and 0.29~mas in right ascension and declination, respectively, are 
significantly larger than the formal errors in the Gaussian 
fitting of the maser features, 0.03-0.1~mas, implying that some systematic errors 
significantly affect our astrometry.
Although it is difficult to estimate the sources of the systematic errors 
in the VLBI astrometry quantitatively (\cite{honma2007}; \cite{hirota2007}), 
they are mainly due to (1) the difference in the optical path lengths 
between the target and reference sources caused by the atmospheric zenith delay residual 
and/or (2) the variability of the structure of the maser feature for the following 
three reasons. 

First, the optical path length error due to an atmospheric zenith delay 
residual of 3~cm, which is typical for VERA observations 
(\cite{honma2007}), 
would cause a relative position error of 0.3~mas, for the case of SVS~13 and J0336$+$3218 
with a separation angle of $1.89^{\circ}$ at an elevation angle of 20~degrees. 
This is comparable to the standard deviation of the post-fit residuals,
especially in declination. 
Because the elevation angle of the sources was always higher than 20~degrees, 
the estimated position error of 0.3~mas gives an upper limit. 
Nevertheless, the atmospheric zenith delay residual could  
contribute significantly to the error sources in our astrometry, 
especially in declination. 

Second, the internal structures in the maser features are found to be 
0.04-0.27~mas based on our channel maps of the maser spots. 
Sometimes, the positions of the maser spots within a feature are spread over more
than 1~mas, as found in the session 2005/021. 
Thus structural changes might explain the 
1~mas declination residual in the data point of session 2004/355 in Figure \ref{fig-ra}. 
This data point is clearly an outlier.   However,
the magnitude of the possible change in the feature positions due to the 
variation of the internal structure of the maser spots certainly could 
be comparable to the dispersion of the positions of the maser spots.
This effect has also been found in the previous results with VERA, in particular 
for the H$_{2}$O maser sources associated with nearby molecular clouds 
(\cite{hirota2007}; \cite{imai2007}). 

Third, the uncertainties in the station positions, delay model, 
and path length errors due to ionosphere would have negligible effects on 
our astrometry, according to the discussions in \citet{honma2007}. 
The uncertainties in the absolute position of J0336$+$3218 are reported 
to be 0.46~mas and 0.56~mas in right ascension and declination, 
respectively (\cite{fey2004}). These uncertainties do not affect the derived annual 
parallax and proper motion because they would add only a constant offsets to 
the position of the maser features. 

In summary, we conclude 
that the astrometric accuracy in our observations is mainly limited 
by the atmospheric zenith delay error and/or the structure of the maser features. 

\subsection{Origin of the H$_{2}$O masers associated with SVS~13}

\begin{figure*}[th]
  \begin{center}
    \FigureFile(160m,160mm){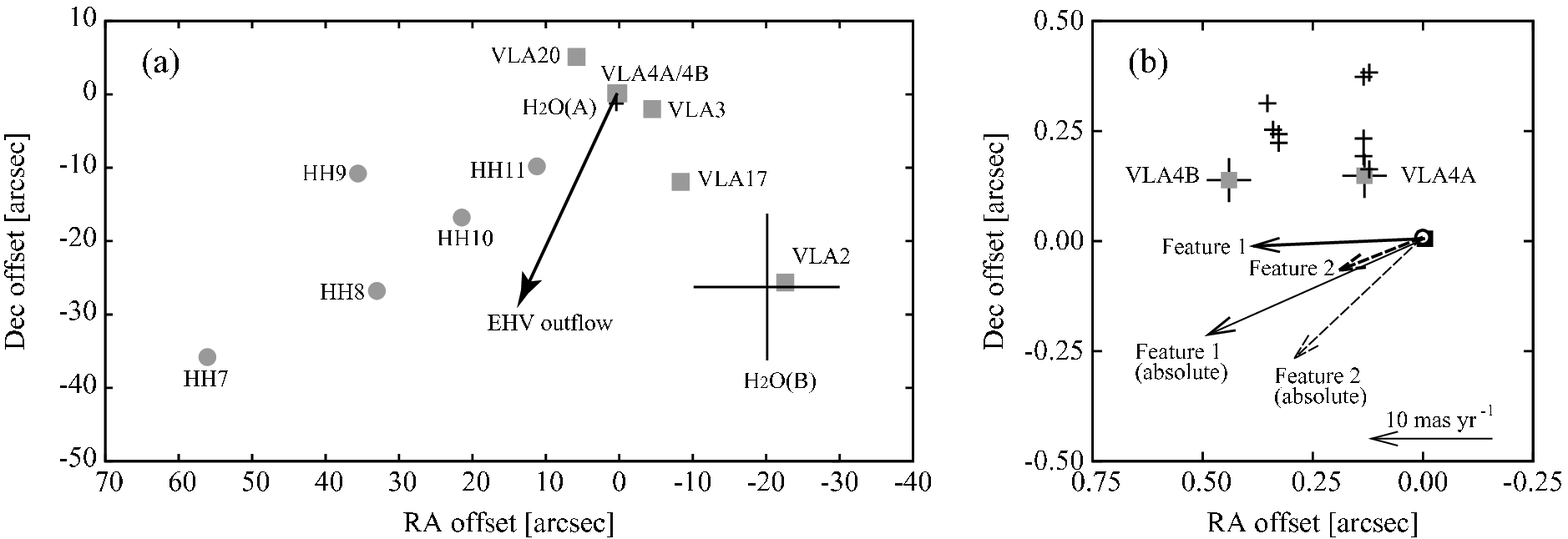}
  \caption{Distribution of H$_{2}$O maser features associated with SVS~13 in NGC~1333. 
   The reference position is the same as in Figure \ref{fig-allspot}.
   The H$_{2}$O maser features detected in our observations are plotted only in the panel (b). 
   (a) Grey squares indicate the positions of the radio continuum sources 
   (\cite{rodriguez1999}; \cite{anglada2000}; \cite{anglada2004}). 
   Grey circles indicate the positions of HH objects (\cite{noriega2001}). 
   Crosses represent the positions of H$_{2}$O masers 
   denoted by H$_{2}$O(A) and H$_{2}$O(B) in \citet{haschick1980}. 
   The size of each cross represents the error bar (1\arcsec \ and 10\arcsec \ for 
   H$_{2}$O(A) and H$_{2}$O(B), respectively). 
   The direction of the EHV outflow is indicated by a bold arrow (\cite{bachiller2000}). 
   (b) Close-up view of the panel (a). Grey squares with error bars (50~mas) indicate 
   the positions of VLA4A and 4B (\cite{anglada2004}). Small crosses 
   represent the positions and error bars (20~mas) of the H$_{2}$O masers detected 
   with the VLA (\cite{rodriguez2002}). 
   An open circle and filled square indicate the positions of 
   the H$_{2}$O maser features 1 and 2, respectively. 
   The absolute proper motion vectors (with respect to the Sun) 
   for the H$_{2}$O maser features 1 and 2 
   are shown with thin solid and dashed arrows, respectively, 
   while those with respect to the LSR are shown with bold solid and dashed arrows. }
  \label{fig-map}
  \end{center}
\end{figure*}

Along with the annual parallax, we successfully measured the absolute positions and 
proper motions of the H$_{2}$O maser features with VERA. As shown in 
Figure \ref{fig-map}(b), the detected maser features are likely to be 
physically related to VLA4A, only 0.2\arcsec \ (50~AU at the inferred distance of 235~pc) 
from the maser features, rather than VLA4B or other sources. The association of 
the maser features in the LSR velocity of 7-8~km~s$^{-1}$ with VLA4A is 
in agreement with the suggestion by \citet{rodriguez2002}. 
They also reported that another group of the H$_{2}$O masers at the LSR velocities 
ranging from $-15$ to $-25$~km~s$^{-1}$ is associated with VLA4B, suggesting 
the presence of an outflow from VLA4B (\cite{rodriguez2002}). 
Although the outflow activity of VLA4B cannot be excluded by our results without 
any detection of the H$_{2}$O masers associated with VLA4B, the present results 
show that VLA4A is the most favorable candidate for the powering source of 
the observed H$_{2}$O masers.  

Figure \ref{fig-map}(b) and Table \ref{tab-results} show the absolute proper 
motions of the maser features associated with SVS~13. 
Both features are moving toward southeast with position angles of 
114$^{\circ}$ and 133$^{\circ}$, respectively, which are apparently in good agreement 
with the alignment of the HH objects traced by the optical emission (\cite{noriega2001}) 
and molecular outflows traced by the CO $J$=2-1 line (\cite{bachiller2000}). 

Although the absolute proper motions of the maser features seem to be well 
aligned with the jets and outflows previously reported, 
we must stress that the proper motions obtained with VERA do not represent 
the ``intrinsic'' proper motions of the maser features because they are measured with 
respect to the Sun and, hence, include the contribution of the Solar motion. 
If we assume the Solar motion relative to the LSR based on the Hipparcos satellite data, 
($U_{0}, V_{0}, W_{0}$)=(10.00, 5.25, 7.17)~km~s$^{-1}$ (\cite{dehnen1998}), 
we can calculate the contribution of the Solar motion to the observed absolute proper motion 
to be 3.7~mas~yr$^{-1}$ and $-7.3$~mas~yr$^{-1}$ in right ascension and declination, respectively. 
Subtracting these values from the observed proper motions listed in Table \ref{tab-results}, 
the proper motions of the H$_{2}$O maser features with respect to the LSR 
in right ascension and declination are obtained to be (14.2, -0.6)~mas~yr$^{-1}$ and 
(6.9, $-2.7$)~mas~yr$^{-1}$, for the features 1 and 2, respectively, as shown 
in Figure \ref{fig-map}(b). They are significantly different from those based on 
the absolute proper motions. 

Using the proper motions with respect to the LSR rather than the absolute ones, 
we will discuss the possible origin of the H$_{2}$O masers associated with SVS~13. 
First, we rule out the possibility that Galactic rotation  
accounts for the observed proper motions, because the contribution of Galactic rotation 
would be only $(-0.65, 0.49)$~mas~yr$^{-1}$, which is estimated by assuming 
$R_{0}$ of 8.0~kpc (\cite{reid1993}) and $\Theta_{0}$ of 236~km~s$^{-1}$ (\cite{reid2004}). 

One of the most plausible explanations is that the proper motions of 
the maser features are due to the jet from VLA4A. 
In fact, VLA4A is proposed to be a powering source of the HH jets and molecular 
outflows as mentioned above (e.g. \cite{noriega2001}; \cite{bachiller2000}).  
If VLA4A is at rest with respect to the LSR, the proper motions of the H$_{2}$O 
masers with respect to the LSR, 
15.8~km~s$^{-1}$ (14.2~mas~yr$^{-1}$) and 8.3~km~s$^{-1}$ (7.4~mas~yr$^{-1}$) 
for the features 1 and 2, respectively, 
represent the motions relative to VLA4A. 
These observed transverse velocities are consistent with the jets and outflows previously 
observed in this region (\cite{wootten2002}). 
However, the directions of the proper motions are not parallel to the jets 
from VLA4A (e.g. \cite{noriega2001}; \cite{bachiller2000}; \cite{wootten2002}). 
In addition, positions of the maser features are not exactly aligned 
with the HH objects and molecular outflows from VLA4A. 
However, VLA4A itself might hava a proper motion with respect to the LSR, 
which could account for the apparent discrepancy found in the geometry of 
VLA4A and jets traced by the H$_{2}$O masers. 

It is worth considering that the observed proper motions are due to 
a rotating circumstellar disk of VLA4A. 
According to \citet{anglada2004}, VLA4B, a companion of the close binary system including 
VLA4A, has a circumstellar disk with the size and mass of $<$30~AU and 0.06$M_{\odot}$, 
respectively, while the mass of the circumstellar disk of VLA4A is at least 5 times 
smaller than that of VLA4B. 
Note, however, that we detected a signature of acceleration of the maser features as 
shown in Figure \ref{fig-vshift}, which might be indicative of rotating disk 
(e.g. \cite{miyoshi1995}). 
If we simply assume an edge-on disk with a rotation velocity $v_{rot}$ of 
12~km~s$^{-1}$, which equals to the mean proper motion of the maser features, 
the radius of the disk, $r_{d}$, can be derived from the relationship $r_{d}=v_{rot}^{2}/A$ 
to be 18~AU,  where $A$ is the drift rate of the radial velocity of the maser features, 
1.7~km~s$^{-1}$~yr$^{-1}$. The inferred radius of the disk is in good agreement with 
that of VLA4B ($<$30~AU; \cite{anglada2004}), but it is slightly smaller than 
the separation from VLA4A, $\sim$50~AU. 
As a result, the enclosed mass within the radius of 18~AU is estimated to be 3$M_{\odot}$. 
This value gives lower limit of the mass because we simply assume the inclination 
axis of the disk to be 0$^{\circ}$. 
The mass of the envelope associated with SVS~13 is derived to be of order  
1$M_{\odot}$ based on the interferometric millimeter wave observations of dust continuum 
emission (e.g. \cite{looney2000}). In addition, the luminosity of SVS~13 is as low as 
$\sim$22$L_{\odot}$ (\cite{jennings1987}), suggesting that it should be a low-mass YSO. 
Although these results agree well with the mass of 3$M_{\odot}$ estimated from 
the circumstellar disk model, the position of the disk (the maser features) 
and the YSO (radio continuum source) is inconsistent with the edge-on disk model. 

The results of our astrometry of the maser features are still puzzling 
because the positions and proper motions of the maser features with respect to 
VLA4A do not fully satisfy the requirements for either a jet 
or a rotating circumstellar disk associated with VLA4A. 
One of the reasons for this discrepancy arises from the lack of the absolute 
proper motion measurements of VLA4A and other possible candidates for 
the powering source of the masers. 
The proper motion vectors could be changed significantly 
by subtracting the absolute proper motion vector of VLA4A, 
making us reconsider all the possibilities discussed above. 
Other possibilities such as an orbital motion of the binary system 
consisting of VLA4A and VLA4B (e.g. \cite{anglada2000}, 2004; \cite{rodriguez2002}) 
could be tested if we measure the proper motions of VLA4A and VLA4B. 
Further high-precision astrometry of the maser features together with the possible candidate 
for their powering sources (i.e. VLA4A and other radio continuum sources in NGC1333) 
is essential to solve this problem (e.g. \cite{rodriguez2005}). 

\section{Summary}

We present the results of multi-epoch VLBI astrometric observations of the 22~GHz H$_{2}$O 
masers associated with SVS~13 in the NGC~1333 region with VERA. 
The principal results of this paper are summarized as follows. 

1. We have determined the annual parallax of the H$_{2}$O 
masers associated with SVS~13 to be 4.25$\pm$0.32~mas, 
corresponding to a distance of 235$\pm$18 pc from the Sun. 
Although the inferred distance of 235~pc is consistent with the photometric 
distance of 220~pc estimated by \citet{cernis1990}, our results provide 
the distance to NGC~1333 
with much higher accuracy than the photometric methods. 

2. The standard deviations of the post-fit residuals for the annual parallax and proper motion 
measurements are 0.10~mas and 0.29~mas in right ascension and declination, respectively. 
The astrometric error sources in our observations are discussed, 
and they are attributed to the difference in 
the optical path lengths between the target and reference sources caused by 
the atmospheric zenith delay errors and/or the variability of the structures of 
the maser features. 

3. The absolute positions and proper motions of the H$_{2}$O maser features 
at the LSR velocities of 7-8~km~s$^{-1}$ are derived, revealing that they are most 
likely associated with 
the radio continuum source VLA4A, based on the observation that the projected distances 
of the maser features are only 50~AU from VLA4A. 

4. We considered the possible origin of the observed proper motions of 
the maser features and found it difficult to explain the positions 
and the proper motions of the maser features in terms of either 
the jet or the rotating circumstellar disk associated with VLA4A 
solely based on the present results. 
Further highly precise astrometric observations of the maser features and 
radio continuum sources including VLA4A would be necessary to reveal 
the complex nature found in the HH~7-11 region in NGC~1333. 

\vspace{12pt}
We thank the referee, Dr. Guillem Anglada, for valuable suggestions and 
Dr. Mark J. Reid for his critical reading of our revised manuscript, 
which substantially improved the paper. 
We are grateful to the staff of all the VERA stations 
for their assistance in observations. 
TH is financially supported by Grant-in-Aids from 
the Ministry of Education, Culture, Sports, Science and Technology 
(13640242 and 16540224).

\end{document}